# Performance degradation of superlattice MOSFETs due to scattering in the contacts

Pengyu Long[1], Jun Huang[1], Zhengping Jiang[1], Gerhard Klimeck[1], Mark J. W. Rodwell[2] and Michael Povolotskyi[1]

[1]School of Electrical and Computer Engineering, Network for Computational Nanotechnology, Purdue University, West Lafayette, Indiana 47907, USA

[2]Department of Electrical and Computer Engineering, University of California, Santa Barbara, California 93106, USA

Ideal, completely coherent quantum transport calculations had predicted that superlattice MOSFETs may offer steep subthreshold swing performance below 60mV/dec to around 39mV/dec. However, the high carrier density in the superlattice source suggest that scattering may significantly degrade the ideal device performance. Such effects of electron scattering and decoherence in the contacts of superlattice MOSFETs are examined through a multiscale quantum transport model developed in NEMO5. This model couples NEGF-based quantum ballistic transport in the channel to a quantum mechanical density of states dominated reservoir, which is thermalized through strong scattering with local quasi-Fermi levels determined by drift-diffusion transport. The simulations show that scattering increases the electron transmission in the nominally forbidden minigap therefore degrading the subthreshold swing ($S.S.$) and the ON/OFF DC current ratio. This degradation varies with both the scattering rate and the length of the scattering dominated regions. Different superlattice MOSFET designs are explored to mitigate the effects of such deleterious scattering. Specifically, shortening the spacer region between the superlattice and the channel from 3.5 nm to 0 nm improves the simulated $S.S.$ from 51mV/dec. to 40mV/dec.

## I. INTRODUCTION

Very-large-scale integration (VLSI) devices require low switching energy ($CV_{DD}^2/2$), low OFF-current ($I_{OFF}$), and large ON-current ($I_{ON}$) for small delay ($CV_{DD}/I_{ON}$). "Simple" supply voltage $V_{DD}$ reduction has dominated the development in Moore's law device scaling to reduce the dynamic power consumption. However, simple $V_{DD}$ reduction increases the OFF current of the devices exponentially and today's devices consume as much power when they are switched or when they hold their state[1]. With the fundamental limit of the turn-on slope of the best possible MOSFET is 60mV/dec Voltage scaling has come to an impasse. Transistors having subthreshold swing ($S.S.$)[2] smaller than the 60mV/dec limit of conventional MOSFETs have therefore attracted great interests. In tunnel FETs, the source valence band filters the injected electron energy distribution, reducing the *S.S.*; unfortunately, $I_{ON}$ is reduced by the low PN junction tunneling probability[3]. In superlattice MOSFETs[4], the source superlattice minigap similarly filters the injected electron energy distribution, again reducing the *S.S.*; because the transmission within the superlattice miniband can approach 100%, $I_{ON}$ can in principle be large, approaching that of conventional MOSFETs.

Reported simulations of superlattice MOSFETs predicted appealing device performances with SS values ranging from 10 to 39mV/dec[4,5,6] using coherent quantum transport models. In these models electrons travel through a quantum mechanical defined miniband coherently, even though the superlattice-shaped source contains a very high carrier density of the order of $10^{18}$-$10^{19}$/cm$^3$. With such high carrier densities it must be assumed that carrier scattering and decoherence effects must be present possibly degrading the perceived advantages of the superlattice injector. Electron-phonon scattering has been shown to reduce the drain current in conventional nanowire MOSFETs[7]. In superlattice MOSFETs, strong scattering in the contacts may also drive electrons towards a thermal energy distribution[8], degrading the superlattice energy-filtering functionality and thereby increasing the S.S. and $I_{OFF}$. To assess the potential application of superlattice MOSFETs in future VLSI circuits, the effect of scattering should therefore be studied to gauge the technical viability of these complex new superlattice MOSFETs.

The self-consistent Born approximation (SCBA) within non-equilibrium Green's function (NEGF) method[9] treats individual scattering processes explicitly. Scattering interactions with various phonons, interface roughness, and alloy disorder have been shown to be strong in resonant tunneling diodes only at low temperature[9,11]. However, the explicit treatment of strong, close to thermalizing electron-electron scattering with an explicit NEGF self-energy has been elusive. Moreover, the method is one or two orders of magnitude more computationally intensive than ballistic simulations[7]. Thermalizing scattering in quantized and continuum injector states in resonant tunneling diodes has, however, been shown to be critical to the quantitative device modeling and prediction[10,11]. This generalized treatment of complex states occupied by a thermalized carrier distribution has been implemented in 1D, 2D, and 3D geometries in NEMO5. The single scattering rate embedded in the generalized contact method represents the effects of multiple scattering mechanisms[12] and can be directly related to experimentally relevant parameters such as the mean free path and momentum relaxation time. This thermalized contact treatment requires the computation of the diagonal elements of the contact Green function using the Recursive Green Function (RGF) algorithm. The thermalized contacts have well-established quasi-Fermi levels determined by solving the drift-diffusion equation[13, 14] in them. For 2D and 3D devices this approach is only about five times slower than the ballistic Quantum Transmitting Boundary Method (QTBM)[15], which, cannot handle such complex, scattering influenced contacts.

This paper first briefly introduces the phenomenological scattering model, and then applies it to study the effects of scattering on the OFF-state (thermalization and broadening) and ON-state characteristics of the superlattice MOSFETs. The effect of spacer on the OFF-state performance and how it can be designed to improve the device S.S are discussed.

## II. DEVICE STRUCTURE AND MODELING METHOD

The double-gate ultra-thin-body (UTB) superlattice MOSFET considered here is an $In_{0.53}Ga_{0.47}As$ channel MOSFET with an $In_{0.53}Ga_{0.47}As/In_{0.52}Al_{0.48}As$ superlattice embedded in its source region. The device structure is illustrated in Fig. 1, with its optimized design parameters listed in Table 1.

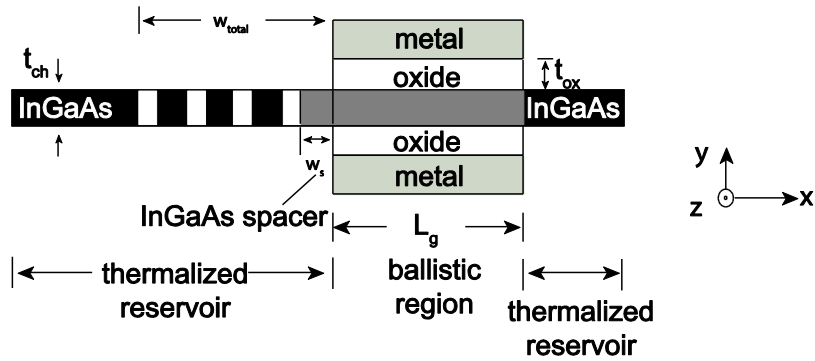

Fig. 1. Device geometry of the superlattice MOSFET. The channel length (thickness) is $L_g$ ($t_{ch}$). The gate oxide thickness is $t_{ox}$, with dielectric constant $\varepsilon_r$. The source (drain) doping density is $N_S$ ($N_D$). The superlattice consists of multiple quantum wells ($In_{0.53}Ga_{0.47}As$) and barriers ($In_{0.52}Al_{0.48}As$) with lengths denoted by $w_i$ and $b_i$. The well doping is $N_w$. The spacer between the superlattice and the channel has length denoted by $w_s$. The structure extends infinitely in the $z$ direction (perpendicular to the page).

TABLE 1. Device parameters of the superlattice MOSFET shown in Fig. 1. Unit ml denotes monolayers.

| Parameter | Value | Parameter | Value |
|---|---|---|---|
| $t_{ox}$ | 2.56nm | $b_1$ | 4ml |
| $\varepsilon_r$ | 20 | $w_1$ | 12ml |
| $t_{ch}$ | 3.2nm | $b_2$ | 6ml |
| $L_g$ | 20nm | $w_2$ | 12ml |
| $N_{S,D}$ | $2.5 \cdot 10^{19} cm^{-3}$ | $b_3$ | 6ml |
| $N_w$ | $1 \cdot 10^{19} cm^{-3}$ | $w_3$ | 12ml |
| $w_s$ | 3.5nm | $b_4$ | 4ml |
| $w_{total}$ | 20nm | $w_s$ | 12ml |

The device is simulated using the atomistic nanoelectronics modeling tool set NEMO5[16,17] by self-consistently solving Poisson's equation and a set of multiscale transport equations. In the multiscale transport approach, the phenomenological scattering model is applied to the source thermalized reservoir, where the carrier energy levels are broadened, and imposed in the drain thermalized reservoir, where carriers dissipate their remaining kinetic energy after ballistic transmission through the central non-equilibrium device. Scattering is neglected in the channel. The separation of scattering free and strong scattering device regions is supported by simulations which showed that the scattering rate in the channel is more than one order of magnitude smaller than that in the contacts[18].

A full non-equilibrium quantum statistical mechanics solution of the transport problem including all relevant particle-particle interactions is in principle possible with the NEGF (Non-Equilibrium Green's Function) formalism. Electron interactions with phonons, interface roughness, or alloy disorder have been treated within NEGF approach in a physically predictive way[19]. Electron-electron scattering, however, is in principle non-local in physical space and energy / momentum space resulting in completely full matrices, which are too large to be solved / inverted for realistically extended devices. No "exact" and practical approach to e-e scattering in NEGF is known at this time. e-e interactions combined with all the other scattering mechanisms are however critically important in the thermalization process in high electron density device regions. Efforts to model high intensity scattering in high electron densities have in principle failed to properly treat the close-to-thermalized electron occupancy in devices such as resonant tunneling diodes, resulting in unphysical descriptions / prediction of bi-stability in devices[8,20]. Rather than spending immense amounts of computational resources to obtain a very-close-to equilibrium distribution with a full NEGF approach it is much more strategic and practical to impose local thermal equilibrium on a quantum mechanically defined density of states[10]. Such an approach subdivides the device into multiple regions whose electron occupancy is rather dramatically different. The source and drain with their very large carrier densities are assumed to be in a local thermal equilibrium, while the central device which is typically tunneling or coherent transport dominated can be treated in non-equilibrium. The equilibrium reservoirs are thermalized with a scattering rate. A physics based scattering rate should in principle be dependent on the distribution of electrons in energy, momentum, and space. Such space, energy, and momentum dependent scattering rates can in principle be computed, but add significant complexity. As a first implementation in NEMO5 we have chosen a scattering rate model that is rather simple: a) constant in energy above the band edge and exponentially decaying under the band edge. The scattering potential is gleened from experimentally observed scattering rates. Spatial and energy variations of the scattering rate within the reservoirs are neglected. We emphasize here that the extended states in the source, not just the local DOS at the end of the source are critical to the injection into the channel.

To summarize the phenomenological scattering model, in the thermalized reservoirs an energy and momentum dependent imaginary potential $i\eta$ is added to the on-site elements of the reservoir Hamiltonian to account for the scattering-induced broadening,

$$\left[G^R(k_z, E)^{-1}\right]_{r,r} = E - H(k_z)_{r,r} - qV_{r,r} + i\eta \tag{1}$$

where $[G^R]^{-1}_{r,r}$ represents the on-site elements of the inverse of the Green's function at position $r$, $k_z$ is the transverse wave vector, $E$ is the energy, $H_{r,r}$ is the diagonal element of the Hamiltonian, $V_{r,r}$ is the potential at position $r$ and $i\eta$ is a small imaginary potential related to the scattering rate. The recursive Green's function (RGF) algorithm[21] is used to obtain in a targeted fashion only the needed matrix elements of the inverted matrix efficiently[9].

The electrons in the reservoirs are assumed to be in equilibrium with local quasi-Fermi levels, given by

$$n_r^{res} = \int \frac{dk_z}{2\pi} \int \frac{dE}{2\pi} A(k_z, E)_{r,r} f_{FD}(E - E_{Fr}) \tag{2}$$

where $f_{FD}$ is the Fermi-Dirac distribution function, $E_{Fr}$ is the local quasi-Fermi level, and $A_{r,r}$ is the $E$ and $k_z$ dependent spectral function in the reservoirs,

$$A(k_z, E)_{r,r} = i\left[G^R(k_z, E)_{r,r} - G^R(k_z, E)^+_{r,r}\right] \tag{3}$$

The quasi-Fermi levels $E_{Fr}$ are determined by solving continuity equation in the reservoirs

$$\nabla \cdot J_r = 0 \tag{4}$$

where $J_r = \mu n_r \nabla E_{Fr}$ is the drift-diffusion current density in the reservoirs, with $\mu$ being the carrier mobility and $n_r$ given in (2). Boundary condition $E_{Fr}=E_{FS}$ ($E_{FD}$) is employed at the left (right) edge of the source (drain) reservoir, where $E_{Fr}=E_{FS}$ ($E_{FD}$) is the Fermi level of the source (drain) contact. We emphasize here that this form of the drift diffusion equation can contain complex, quantum mechanically defined local charge densities. The current continuity throughout the whole device is established by enforcing the current at the reservoir-channel interfaces to be equal to the ballistic current of the central region[13]

$$I_{LR} = \int \frac{dk_z}{2\pi} \int \frac{dE}{2\pi} T(k_z, E)[f_{FD}(E - E_{FL}) - f_{FD}(E - E_{FR})] \tag{5}$$

where $T$ is $E$ and $k_z$ dependent transmission coefficient, $E_{FL}$ and $E_{FR}$ are the quasi-Fermi levels at the left and right boundaries of the channel region.

A Newton solver is developed to solve the non-linear equation (4). Once the quasi-Fermi levels are determined, the electron density distribution in the reservoirs is obtained by (2) and the density distribution in the channel is obtained by

$$n_r^{ch} = \int \frac{dk_z}{2\pi} \int \frac{dE}{2\pi} [A_L(k_z, E)_{r,r} f_{FD}(E - E_{FL}) + A_R(k_z, E)_{r,r} f_{FD}(E - E_{FR})] \tag{6}$$

where $A_L$ and $A_R$ are the left-connected and right-connected spectral functions in the channel. Figure 2. depicts the process for the self-consistent solution to the Poisson equation with the multi-scale transport.

The simulations use scattering potentials $\eta$ of 5meV and 10meV, corresponding to scattering time $\tau$ of 66fs and 33fs, respectively, similar to values from the literature[22]. Khondker[23] *et al.* considered phonon and random impurity scattering and concluded that when the resonant states are filled, $\tau$ is approximately 80fs. Given a $\sim 1.5 \cdot 10^7$ cm/s thermal velocity $v_{th}$, the mean free path $\lambda = v_{th}\tau$ is 10.5nm for $\eta$ =5meV and is 5.3nm for $\eta$ =10meV. From $\mu = q\tau/m^*$, where m$^*$ is the electron effective mass of confined In$_{0.53}$Ga$_{0.47}$As. $\eta$ =5meV corresponds to $\mu$ of 1206 cm$^2$/(V·s) and 10meV corresponds to 603 cm$^2$/(V·s).

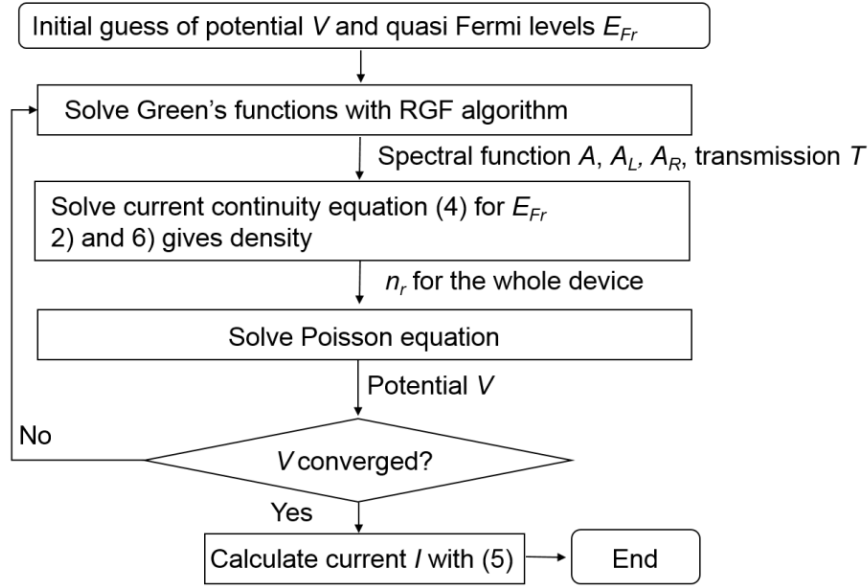

Fig. 2. Flowchart of the multi-scale transport simulation with ballistic transport in the channel and phenomenological scattering model in the reservoirs.

## III. DISCUSSION

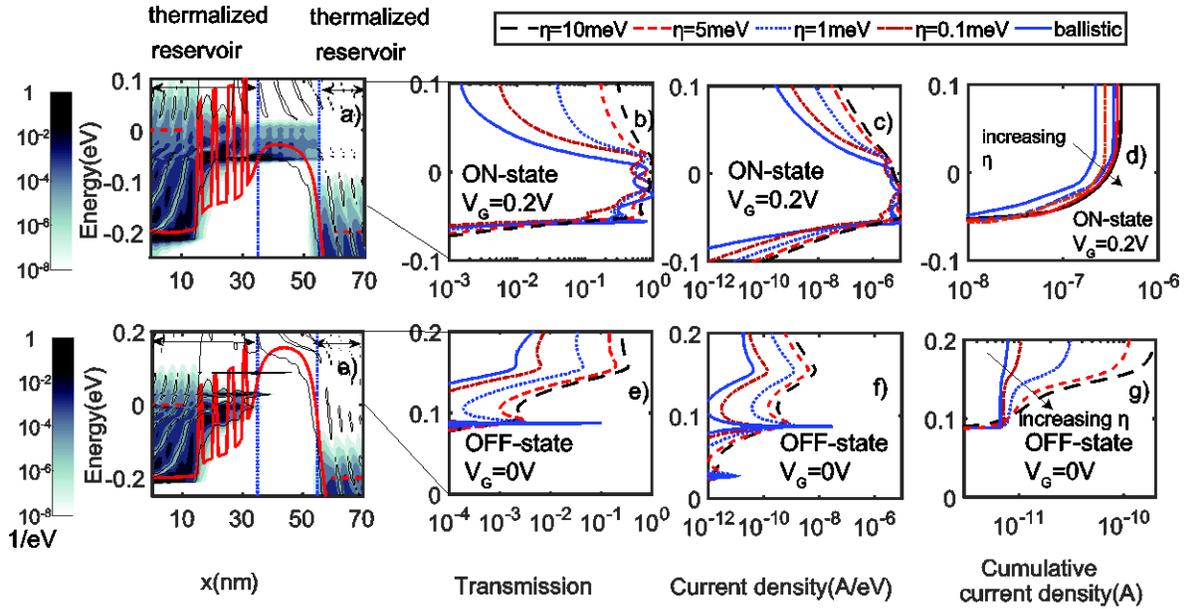

Fig.3 a) Energy band diagram and electron density profile of the superlattice MOSFET in an ON-state bias with the scattering and ballistic regions indicated; b) transmission probability versus energy given a zero transverse wave vector $k_z$; c) energy-resolved current density $J(E)$ d) cumulative current density for $k_z=0$. As η increases, the transmission in the off-resonant minigap increased exponentially, and resonant states are broadened. In ON current increases slightly by about a factor of 2. e), f), g) & h) plot the energy band diagram, transmission, energy-resolved current, cumulative current in an OFF-state state bias. In plots b), c), d), f), g) and h), imaginary scattering potentials η of 0.1meV, 1meV, 5meV, and 10meV are applied. As η increases, the transmission in the off-resonant minigap increased exponentially, and the resonant states are broadened. The OFF current increases exponentially by by a factor of 25x. The gate voltage is NOT scaled for $I_{OFF}$ normalization.

The effect of scattering in the reservoirs on the transmission probability through the central device is studied by varying the imaginary scattering potential. The transistor off-state is defined with $V_{DS} = 0.2$V, $V_{GS} = 0$V, and with the threshold voltage adjusted so that the off-state current, given zero scattering, is 0.1A/m. The transistor ON-state is defined with $V_{DS} = V_{GS} = 0.2$V.

Fig. 3 shows the band diagram, transmission, and energy-resolved current density for the OFF- and ON-states (top and bottom row, respectively). The effect of scattering in the source is examined by ramping the empirical scattering potential from 0.1meV to 10meV. ON and OFF current show different sensitivities to the increased rate of scattering in the source.

Fig. 3a depicts the local density of states superposed with the local charge density, in addition to the bulk-based bandedges and the Fermi Levels in the ON state. The injection of carriers from the superlattice emitter into the channel region is clearly visible. The electrons travel ballistically through the channel region and are, by the model reservoir assumption, rapidly thermalized in the collector/drain.

Fig. 3b. zooms into the energy range where the miniband injection occurs and plots the transmission coefficient. As the empirical scattering rate is increased, the energetic definition of the miniband is softened and the ripples in the transmission window are softened a little bit. Fig. 3c plots the corresponding energy dependent current density. Given the high bias between source and drain this current density is basically the transmission in Fig. 3b multiplied by the Fermi function in the injection site. Due to the exponential decay of the Fermi function the dramatic exponential broadening of the transmission is counteracted. From Fig. 3c, however, it is hard to tell if the current will increase or decrease. It is instructive to define the following quantity which we entitle"running integral of the current density". This integral converts the resonance dominated current channels into step-like contributions to the total current.

$$J_{int}(E,k) = \int_{E_0}^{E} dE' \, J(E',k)$$

Fig. 3d shows that the overall current increases slightly as the scattering strength is increased. However, in the ON state, both with weak and strong scattering, current is primarily carried by the first superlattice miniband. The overall current is increased by a 86% as eta is increased to a maximum value of 10meV.

The lower row of Fig. 3 compares the OFF state behavior to the top row ON state behavior. In the OFF state the miniband is nominally blocked by the gate potential and remains reasonably unaffected by the scattering strength. The off-resonance transmission *above* the miniband, however, is significantly increased with increased scattering strength as clearly visible in Fig. 3e. The current density panel in. 3f shows that there is indeed significant current flowing at high energies, above the miniband. The cumulative current density in Fig. 3g shows clearly that the OFF current increases exponentially by a factor of 25.

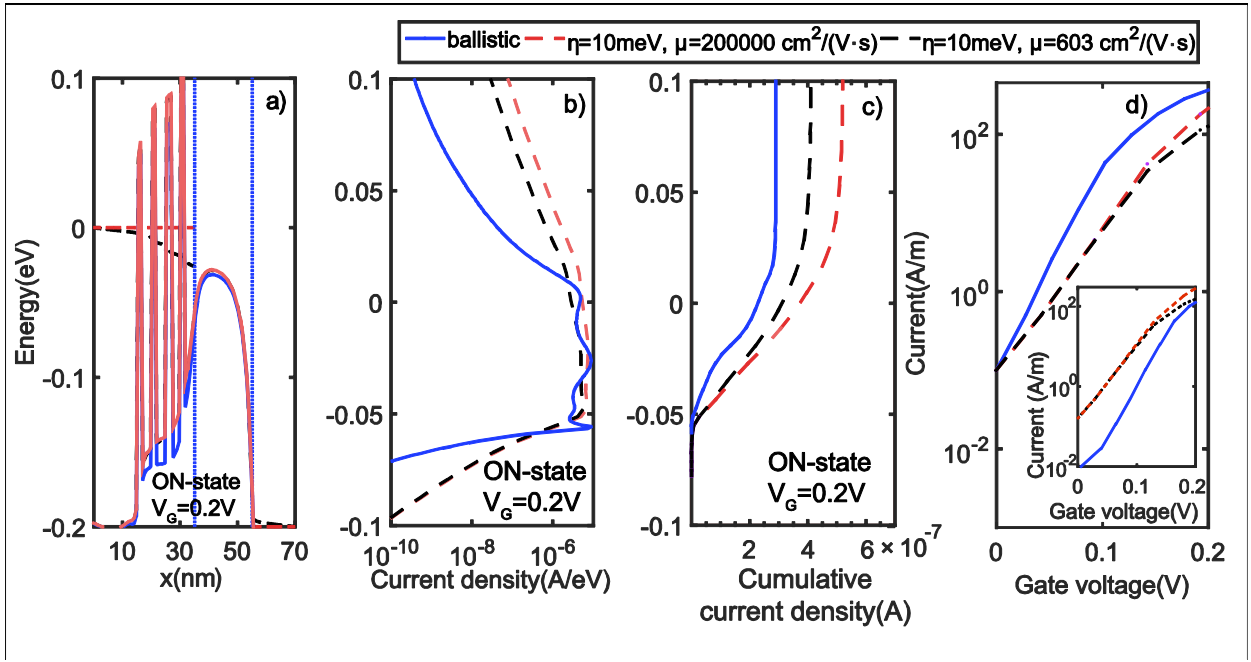

Fig. 4 a) ON-State band diagram and quasi-Fermi level with mobility $\mu=20000$ cm$^2$/(V·s) and $\mu=603$ cm$^2$/(V·s). b) Energy resolved current density $J(E)$ for $k_z=0$ at ON-state. c) Cumulative current density d) Transfer characteristics of superlattice MOSFET calculated with different values of mobility $\mu$, comparing with ballistic simulation aligned at $I_{OFF}=0.1$A/m. Inset in Fig 4d) shows transfer characteristics without this alignement. A smaller mobility causes the quasi-Fermi level to decrease more across the superlattice resulting in an $I_{ON}$ decrease, further degrading overall ON/OFF ratio. Other than d), the gate voltage is NOT scaled for $I_{OFF}$ normalization.

Another effect of scattering in the emitter is a resistive transport behavior, which results in a Fermi level drop, and a self-consistent deformation of the semiconductor band edge. A Fermi level and potential drop in the source in turn change the superlattice alignment and change the injection into the central device. Sophisticated drift diffusion models with various scattering mechanisms are available today in commercial simulators [24,25,26] for standard semiconductor devices. These simulators do in general do not handle heterostructures within a quantum mechanical approach that details a spatially, and energy and momentum dependent scattering. Mobility models for such heterostructures are subject to commercial development. Here we are limited to the empirical exploration of simple mobility models that involve an "exact" density of states. We chose to explore the effects of a finite empirical mobility coupled to the Drift Diffusion equation with mobility $\mu$ of 200,000 cm$^2$/(V·s), and the other with $\mu$ of 603 cm$^2$/(V·s). As the mobility decreases, the quasi-Fermi level drops more dramatically across the superlattice and the drain reservoirs. As a result, the transmission window across the channel is reduced causing the ON-current to decrease from 230A/m to 160A/m with a normalized $I_{OFF}$=0.1A/m. The insert of 4d shows the I-V curves without OFF current normalization. Clearly the OFF current increases significantly with scattering (factor of 17), while the ON current increases slightly (factor of 2.15). The introduction of a low mobility in the presence of scattering broadening hardly affects the OFF current (1% lower) but lowers the ON current by 75%.

In reality, the mobility is related to the mean free time and thus scattering rate. To reflect this relation, the mobility in the rest of this paper will be determined from $\mu = q\tau/m^*$, where m$^*$ is the electron effective mass of confined In$_{0.53}$Ga$_{0.47}$As. Mean relaxation time $\tau = \hbar/(2\eta)$. $\eta$ =5meV corresponds to $\mu$ of 1206 cm$^2$/(V·s) and 10meV corresponds to 603 cm$^2$/(V·s).

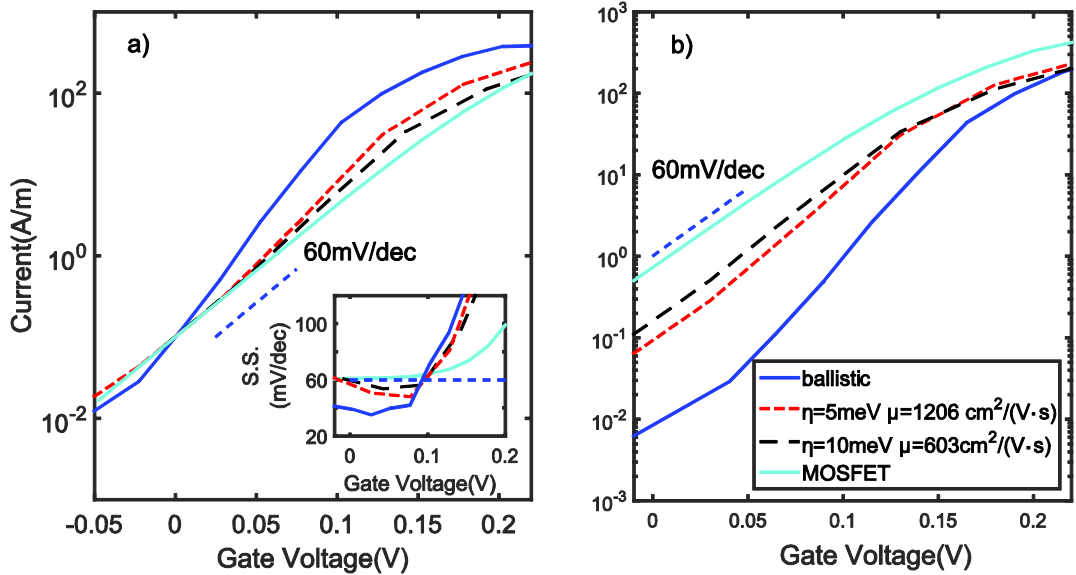

Fig. 5 Transfer characteristics of the superlattice MOSFET calculated with different values of imaginary scattering potential (and mobility), in comparison with the conventional MOSFET with identical structure except for the superlattice. a) I-V data is aligned at the same $I_{OFF}$ of 0.1A/m by adjusting metal workfunction of the gate. b) Without adjusting the metal workfunction (no $I_{OFF}$ normalization). The simulations predict an exponential increase in $I_{OFF}$ and degraded S.S. with the introduction of scattering.

Fig. 5 shows the $I_D$-$V_G$ characteristics with the threshold now adjusted to normalize $I_{OFF}$ to 0.1A/m for all devices. The S.S. (Fig. 5 a)) degrades from 39mV/dec. in the ballistic case to 51mV/dec at $\eta$ = 5meV and 60mV/dec at $\eta$ =

10meV. The ON-current degrades from 360A/m in the ballistic case to 190A/m at $\eta = 5$meV and 130A/m at $\eta = 10$meV. The degradation in ON/OFF ratio is partially caused by the increase in OFF-state leakage current and partially caused by the quasi-Fermi level drop in the thermalized reservoirs. The insert in Fig. 5a shows the value of the S.S. as a function of gate voltage. On this voltage scale the region of sub 60mv/dec is below 0.1V.

Fig. 5b shows the same data of Fig. 5a without the $I_{OFF}$ normalization. The increase in the OFF current is visible on the exponential scale. On this voltage reference the ON current is hardly changed by the scattering.

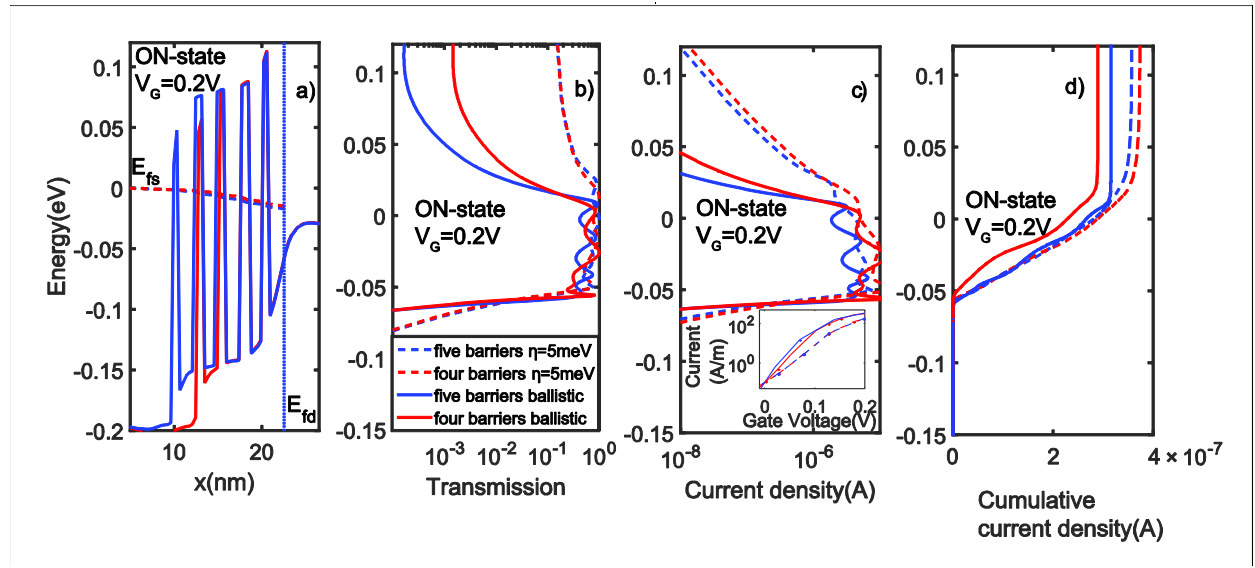

Fig. 6. Exploration of two different superlattice lengths with four and five barriers. (a) Band diagrams and quasi Fermi levels at ON-state. b), c) &d) Corresponding transmission probabilities, energy resolved current, cumulative current for superlattices with four and five barriers, in the ballistic limit and with scattering in the contacts ($\eta = 5$meV). Inset in Fig. 6c) shows ballistic and scattering transfer characteristics of four and five barrier superlattices. In the ballistic transport case, the addition of barriers refines the transmission through the miniband as expected. However, broadening due to incoherent scattering counteracts the definition of the minibands and raises the non-resonant band tails, which ultimately increases the OFF current and results in similar S.S. with four and five barrier superlattices. The gate voltage is NOT scaled for $I_{OFF}$ normalization.

Figure 6 depicts the effects of scattering in superlattices of different lengths (four and five barriers). The transmission probability is plotted in the transistor ON state, both in the ballistic case and with $\eta = 5$meV, for superlattices having four and five barriers. In the ballistic case, the off-resonant transmission in the superlattice minigap decreases as the number of superlattice periods is increased (Fig. 6a), as expected. The miniband becomes more well-defined.

The 5meV scattering strength corresponds to a decoherence length of 9.9nm, the transmission in the superlattice minigap is similar for the four-barrier and five-barrier superlattice designs, and there is little benefit in using more superlattice periods. In fact on this voltage scale here it appears that while the coherent 5 barrier structure delivers a higher current than the 4 barrier structure, the opposite is true when scattering is present (Fig. 6d).

An additional negative effect due to adding more superlattice cells can be understood intuitively in the context of scattering and thermalization. If there is strong thermalization in the superlattice due to the high carrier densities, then adding more cells begins to drop the Fermi level over a longer leverage arm, stretching out the I-V characteristic over a wider range of source-drain biases, thus reducing the ON current by moving the highest possible current out of the voltage of interest below 0.2V. Indeed this is what Figure 7a) shows. The Gedankenexperiment in Fig. 7 is constructed such that in a nominally unchanged device, the region in which incoherent scattering is enforced is increased from just the left contact to enclose more and more barriers of the superlattice. The voltage scale is not changed between these simulations (no normalization of $I_{OFF}$). As the number of superlattice periods increases, the quasi-Fermi levels decrease more in the thermalized reservoirs, which affects the ON-state current. Therefore, the modeling of incoherent scattering leads to critical device design insight and in this case simplifies the overall design of the device as only a limited number of barriers are needed.

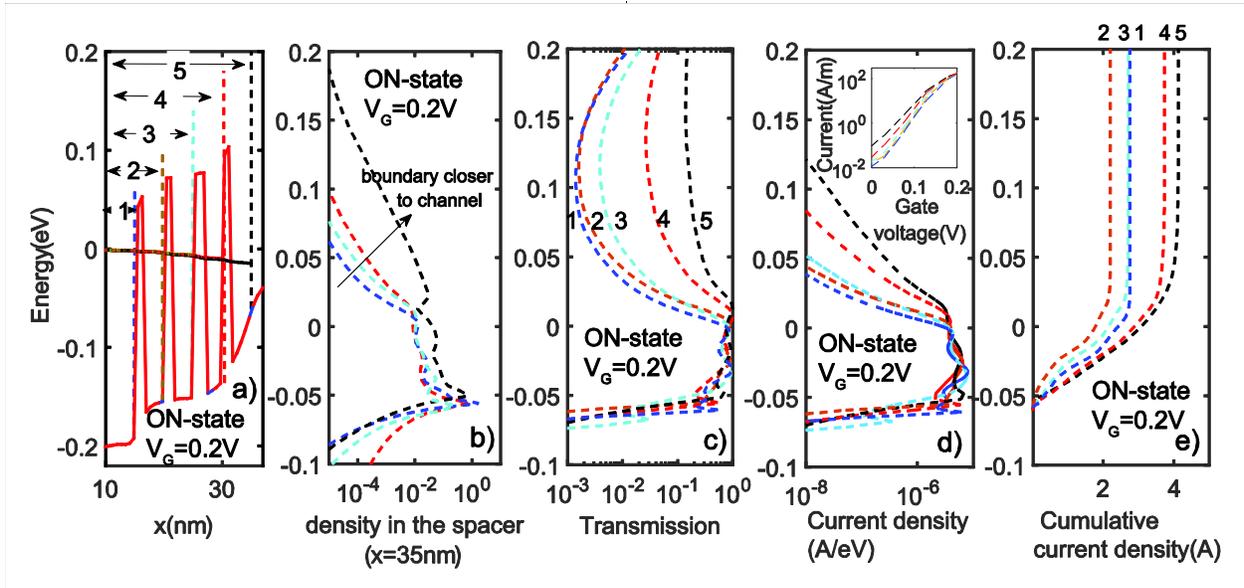

Fig. 7. Energy-resolved dependence analysis of the extent of the reservoir treatment. (a) Band edge diagram in the ON-state at 0.2 V for the completely ballistic device. (b) Electron density closely in front of the channel (x=35nm) (b), transmission probability (c), energy resolved current (d), cumulative current density (e) for different regions where scattering takes place (a) for a four-barrier superlattice. A 5meV scattering potential is present in the indicated regions (1 through 5) and in the drain, with ballistic transport assumed in the rest of the transport path. Inset in Fig. 7d) shows transfer characteristics with different boundaries. The superlattice energy filtering is most rapidly degraded by scattering immediately adjacent to the channel (in the spacer). The gate voltage is NOT scaled for $I_{OFF}$ normalization.

The modeling with NEMO5 can also address the critical issue to determine the region in which scattering has the most detrimental effect. Fig. 7c) shows the energy-dependent superlattice transmission as a function of the extent of the region in which thermalizing scattering is imposed. The superlattice energy filtering is most rapidly degraded by scattering immediately adjacent to the channel (in the spacer); electrons that are only scattered close to the source will still have steep energy-dependent filtering as they pass through the remainder of the superlattice.

The critical dependence on the extent of the scattering region, or from a different point of view the strong dependence of the transport on the few nanometers close to the channel is also illustrated in Fig.7b). If the boundary is set close to the source (boundary 1 in Fig. 7a), transport through the superlattice and channel is ballistic, the electron density is well confined to the energy range of miniband (Fig. 7b). If the boundary is set to left of the channel (boundary 5 in Fig. 7a), the electron density above and below the miniband increased (Fig. 7b). The increase in density is because 1) electrons re-thermalized after crossing the superlattice region, counterweighing the effects of energy filter; 2) the quasi-bound state in the spacer (x=35nm) is broadened. This change in density causes the superlattice miniband to move up slightly in energy, as shown in Fig 7c).

The simulations of Fig. 7 indicate that the superlattice FET subthreshold swing and OFF-current are strongly degraded by scattering in the region between the superlattice and the channel potential barrier. To investigate this, the energy-dependent transport characteristics, and current-voltage characteristics, are simulated comparing a case with spacer and without, both under the condition of scattering parameter $\eta$ setting to 5meV. The voltage scales are scaled to normalize $I_{OFF}$. As the spacer is removed, the transmission in the superlattice minigap becomes smaller (Fig. 8b) and the device $I_D$-$V_{GS}$ characteristics become steeper (Fig. 8e). The subthreshold swing improves from 51mV/dec. with $w_s$ = 3.5nm to 40mV/dec. with $w_s$ = 0nm.

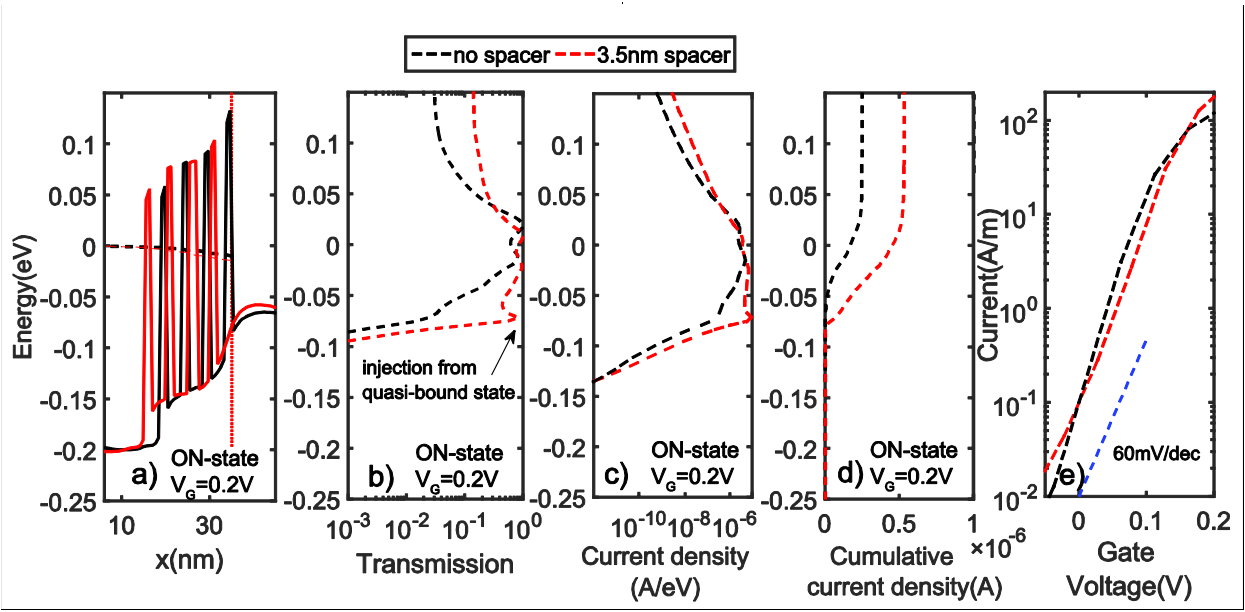

Fig. 8 Exploration of an optimized design that avoids the spacer between the superlattice and the channel assuming a 5meV scattering potential in the source in ON state. a) energy band diagrams including a 3.5nm spacer (red line) and without a spacer between the superlattice and the channel (black line); b) transmission c) energy resolved current density d) cumulative current. As the spacer layer is reduced and superlattice pushed closer to the gate the injection from quasi-bound state diminished, the total current decreased. e) $I_{DS}$-$V_{GS}$ characteristics at $V_{DS}$=0.3V, computed with $\eta$= 5meV, for a 3.5nm spacer and no spacer between the superlattice and the channel. in e) The gate voltage is adjusted to set the OFF current to 0.1 A/m.

As the spacer is removed, the potential profile within the superlattice is more strongly modulated by the gate potential (Fig. 8 (a)). This modulation misaligns the states within the superlattice, perturbing the ON-state transmission characteristics (Fig. 8b). In addition, the injection from the quasi-bound state (inside the spacer) is also removed, causing $I_{ON}$ to decrease (Fig. 8e) (under a condition of constant $I_{OFF}$). This somewhat counter-intuitive result displays the critical interplay of the incoherent effects in the injector superlattice and the transport through the central device region. Addressing these incoherent effects are critical to predicting and understanding performance of today's steep S.S. transistors.

## IV. SUMMARY

Previous work had proposed superlattices to be inserted into the source of a MOSFET to overcome the 60mV/dec limitation. These model predictions assumed a perfectly coherent transport throughout the device region. In this work, the effect of carrier scattering on the DC characteristics of 2D superlattice MOSFETs has been analyzed in detail. This simulation work shows that reasonable assumptions of scattering in the high carrier density source regions degrades the device performance from the coherent theoretical limit. Scattering in the superlattice due to high carrier concentrations degrades the transistor subthreshold swing ($S.S.$) and $I_{ON}/I_{OFF}$ current ratio. Given significant scattering, increasing the number of superlattice filter periods does not significantly improve the transistor performances. These model predictions therefore simplify the overall device design by a quantitative metric. As the device characteristics are most strongly degraded by scattering in the spacer region immediately before the channel, removing the spacer between the superlattice and the channel improves the subthreshold swing, but at the cost of degrading the ON current.

**Acknowledgment**

The use of nanoHUB.org computational resources under NSF Grant Nos. EEC-0228390, EEC-1227110, EEC-0634750, OCI-0438246, OCI-0832623 and OCI-0721680 is gratefully acknowledged. This material is based upon work supported under NSF Grant (1509394). NEMO5 developments were critically supported by an NSF Peta-Apps award OCI-0749140 and by Intel Corp.